\documentclass[aps,prl,twocolumn,superscriptaddress]{revtex4-2}
\usepackage{graphics,graphicx, array, afterpage}
\usepackage{qcircuit,amsmath}
\usepackage{grffile}
\usepackage[export]{adjustbox}
\usepackage{lineno}
\usepackage{xcolor}
\usepackage{longtable}
\usepackage{braket}
\usepackage{array}
\usepackage{multirow}
\usepackage{enumerate}
\usepackage{amsmath}
\usepackage{amssymb}
\usepackage{amsthm}
\usepackage{times}
\usepackage{multirow}

\usepackage[colorlinks, citecolor=blue]{hyperref}
\hypersetup{linkcolor=magenta,urlcolor=blue,citecolor=blue,pdfstartview={FitH},urlcolor=blue}

\begin{document}

\title{Generalized Spin Helix States as Quantum Many-Body Scars in Partially Integrable Models}

\author{He-Ran Wang}
\email{whr21@mails.tsinghua.edu.cn}
\affiliation{Institute for Advanced Study, Tsinghua University, Beijing 100084, People's Republic of China}

\author{Dong Yuan}
\email{yuand21@mails.tsinghua.edu.cn}
\affiliation{Center for Quantum Information, IIIS, Tsinghua University, Beijing 100084, People's Republic of China}

\begin{abstract}
Quantum many-body scars are highly excited eigenstates of non-integrable Hamiltonians which violate the eigenstate thermalization hypothesis and are embedded in a sea of thermal eigenstates.
We provide a general mechanism to construct partially integrable models with arbitrarily large local Hilbert space dimensions, which host exact many-body scars. We introduce designed integrability-breaking terms to several exactly solvable spin chains, whose integrable Hamiltonians are composed of the generators of the Temperley-Lieb algebra.
In the non-integrable subspace of these models, we identify a special kind of product states -- the generalized spin helix states as exact quantum many-body scars, which lie in the common null space of the non-Hermitian generators of the Temperley-Lieb algebra and are annihilated by the integrability-breaking terms.
Our constructions establish an intriguing connection between integrability and quantum many-body scars, meanwhile provide a systematic understanding of scarred Hamiltonians from the perspective of non-Hermitian projectors.

\end{abstract}

\maketitle 

\textit{Introduction.}-- The Hamiltonians of isolated quantum many-body systems typically respect the Eigenstate Thermalization Hypothesis (ETH) \cite{Deutsch1991Quantum,Srednicki1994Chaos,Deutsch2018Eigenstate}, which leads to the quantum thermalization dynamics and scrambling of local information during the unitary evolution. Two mechanisms strongly violating the ETH have been extensively studied in previous literature, including the integrability \cite{Sutherland2004beautiful} and many-body localization \cite{Nandkishore2015Manybody,Abanin2019Colloquium}. In the past few years, a type of weak ergodicity breaking phenomena known as quantum many-body scars (QMBS) has attracted considerable attention \cite{Serbyn2021quantum,Moudgalya2022Quantum,Chandran2022Quantum}. QMBS refers to a vanishing fraction of highly excited eigenstates that violate the ETH and possess relatively low entanglement compared to neighboring thermal eigenstates. The studies of QMBS were initiated by the discovery of nonthermal coherent revival dynamics in Rydberg atom simulators \cite{Bernien2017Probing,Bluvstein2021Controlling}. Since then numerous theoretical works have been devoted to understanding these anomalous excited eigenstates \cite{Turner2018weak,Turner2018quantum,shiraishi2019connection,Khemani2019Signatures,Ho2019periodic,Choi2019emergent,Iadecola2019Quantum,Surace2020Lattice} and finding new models hosting QMBS \cite{Moudgalya2018Exact,Schecter2019Weak,bull2019systematic,ok2019topological,Iadecola2020Quantum,hudomal2020quantum,Desaules2021Proposal,Banerjee2021Quantum,karle2021area,Langlett2022rainbow,Desaules2022Weak}.


Among the plethora of Hamiltonians hosting QMBS, there exist many models in which scarred eigenstates have exact analytical expressions \cite{Moudgalya2018Entanglement,lin2019exact,Lin2020Quantum,Surace2021Exact,Chattopadhyay2020quantum,Moudgalya2020eta,Mark2020eta,Moudgalya2020Large,Shibata2020Onsager,Zhang2023Extracting,Yuan2023Exact,buvca2019Non,Wang2023Embedding}.
These exact QMBS can be characterized by several unified frameworks such as the projector embedding \cite{Shiraishi2017Systematic}, group-theoretic approach \cite{nicholas2020from,Pakrouski2020Many,Pakrouski2021Group,Ren2021Quasisymmetry}, and commutant algebra \cite{Moudgalya2022Exhaustive}. 
In particular, the scar states in Ref.~\cite{Shibata2020Onsager} were constructed by adding integrability-breaking terms to exactly solvable clock models \cite{Vernier2019Onsager}. These terms annihilate a set of special excited eigenstates of the original integrable Hamiltonian. Although the integrability is broken in the presence of perturbations, these special eigenstates are preserved as exact QMBS in the resulting non-integrable Hamiltonian.

\begin{figure}[ht]
\centering
\includegraphics[width=0.9\linewidth]{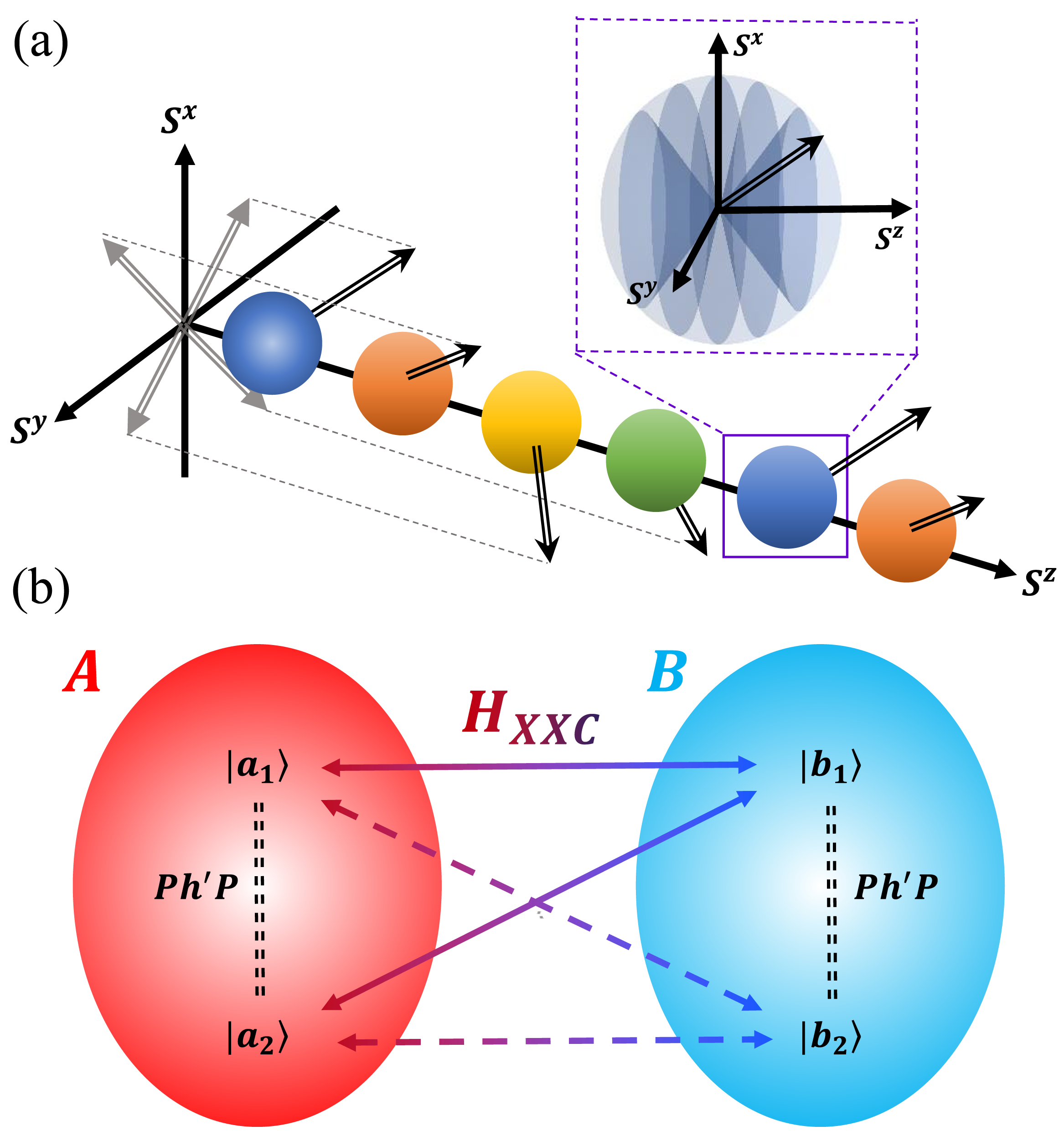} 
\caption{(a) Generalized spin helix states for higher-spin models. The spin vectors (the black arrows) of the higher spins (the colorful spheres) wind in the $S^x–S^y$ plane, while maintaining a fixed angle with the $S^z$ axis. 
(b) A pictorial illustration for the partially integrable Hamiltonian Eq.~\eqref{eq:perturbed} hosting the generalized spin helix states as exact quantum many-body scars. The local degrees of freedom on each site are divided into two sets $A$ and $B$. The interactions between the basis states in $A$ and those in $B$ constitute the XXC Hamiltonian, while the interactions between different basis states within $A$ or $B$ take the form of $Ph'P$ (see the main text).
}
\label{Illustration}
\end{figure}

In this paper, we greatly extend this type of philosophy by providing a general strategy to construct several partially integrable models hosting exact many-body scars.
Specifically, we introduce designed integrability-breaking terms to the higher-spin generalizations of the one-dimensional anisotropic Heisenberg model. These models host a special kind of product states -- the generalized spin helix states [see Fig. \ref{Illustration}(a) for a schematic illustration] as excited eigenstates, which are annihilated by the perturbation terms and thus preserved as exact QMBS in their non-integrable subspaces. To provide some background, we introduce the exactly solvable spin-$1/2$ XXZ Hamiltonian \cite{yang1966oneI}:
\begin{equation*}
    H_{\text{XXZ}}=\sum_{j=1}^L \left[ \sigma_j^x\sigma_{j+1}^x+\sigma_j^y\sigma_{j+1}^y-\cos(\gamma)(\sigma_j^z\sigma_{j+1}^z-1) \right],
\end{equation*}
where $\sigma_j^{\alpha}~(\alpha=x,y,z)$ denote Pauli matrices on the site $j$, and $\cos(\gamma)$ parameterizes the anisotropy of the interaction strength in the $z$ direction. The following spin helix states are zero-energy eigenstates of the above Hamiltonian: $\ket{\psi(\beta)}=\bigotimes_{j=1}^L (\ket{\uparrow}_j+\beta\exp(i(j-1)\gamma)\ket{\downarrow}_j)$, $\beta\in \mathbb{C}$. The term ``spin helix" refers to the spatial spin configuration in which the azimuthal angle of each spin increases by a fixed value $\gamma$ compared to the previous one. These highly excited eigenstates are associated with the phantom Bethe roots with zero energy and transparent scattering amplitudes \cite{pasquier1990common,gainutdinov2016algebraic,Zhang2021Phantom,Popkov2021phantom}. In addition, these states can exhibit intriguing non-equilibrium many-body dynamics: After we couple a uniform $z$ direction Zeeman field to the spin chain, $\ket{\psi(\beta)}$ is no longer an eigenstate. Instead, the parameter $\beta$ oscillates and shows perfect revival, as observed in recent cold-atom experiments \cite{jepsen2020spin,paul2021transverse,jepsen2022long}.

The long-lived non-stationary dynamics starting from the spin helix states explicitly give rise to the violation of generalized thermalization in integrable models, which states that generic initial states typically thermalize to generalized Gibbs ensembles after long-time evolution \cite{Rigol2007Relaxation,Cassidy2011Generalized}. 
Recently, Refs.~\cite{Lee2020Exact,Medenjak2020Isolated,Chertkov2021Motif,jepsen2022long,Tang2022Multimagnon,Shi2023Robust,Zhang2023Generalized,Gerken2023Product} have embedded spin helix states as ``true" QMBS through adding certain integrability-breaking terms to the spin-$1/2$ XXZ Hamiltonian, or extending to higher spatial dimensions. In the current work, we go beyond the spin-$1/2$ cases by allowing arbitrarily large local degrees of freedom on each site, while maintaining the two-local property of the Hamiltonians. 
After adding designed perturbations, the Hilbert space factorizes into several integrable subspaces in which the Hamiltonian reduces to the spin-$1/2$ XXZ Hamiltonian, together with one non-integrable subspace hosting the generalized spin helix states as QMBS. This unique feature makes our models  versatile platforms for investigating the interplay between integrability and weak ergodicity breaking. 
More interestingly, we could interpret our partially integrable Hamiltonian as the summation of \textit{local non-Hermitian projectors} annihilating the generalized spin helix states. We demonstrate that this underlying mechanism also exists in many other scarred Hamiltonians~\cite{Mark2020Unified,nicholas2020from,Moudgalya2022Exhaustive}, which provides another crucial unified framework for exact QMBS. 

\textit{The scarred models with partial integrability.}-- 
Our constructed scarred models consist of two components:
\begin{equation}\label{eq:perturbed}
    H=H_{\text{XXC}}+\sum_{j=1}^L P_{j,j+1}h'_{j} P_{j,j+1}.
\end{equation}
The first component is the integrable XXC model consisting of of translational invariant two-local Hamiltonian densities~\cite{Maassarani1998XXC}:
\begin{align}\label{eq:Multi}
H_{\text{XXC}}=&\sum_{j=1}^L\sum_{a\in A,b\in B} [\eta_{b} E^j_{ab}\otimes E^{j+1}_{ba}+\eta_{b}^{-1} E^j_{ba}\otimes E^{j+1}_{ab}\nonumber\\
&-\cos(\gamma)E^j_{aa}\otimes E^{j+1}_{bb}-\cos(\gamma)E^j_{bb}\otimes E^{j+1}_{aa}].
\end{align}
Here $L$ is the number of sites on a one-dimensional chain with the periodic boundary condition.
For each site, the basis states of the local Hilbert space are labeled by $\ket{c},c=1,2\cdots, N$. The local degrees of freedom are divided into two sets $A$ and $B$, with the dimension $N_A$ and $N_B$ respectively. The twist parameter $\eta_{b}=\pm 1$. $\cos(\gamma)$ parameterizes the interaction strength between basis states belonging to different sets on the neighboring sites.  The two-local Hamiltonian densities $h_{j,j+1}$ are represented in terms of the standard basis elements of $N\times N$ matrix $E^j_{ab}=(\ket{a}\bra{b})_j$. 

The second component in the $Ph'P$ form is introduced to break the integrability. $P_{j,j+1}$ are two-local Hermitian projectors given by
\begin{equation}\label{eq:projector}
P_{j,j+1} =\sum_{a,a'\in A}(P^{+1}_{aa'})_{j,j+1}+\sum_{b,b'\in B}(P^{\eta_b\eta_{b'}}_{bb'})_{j,j+1},
\end{equation}
where 
$P_{ab}^\eta=\frac{1}{2}(\ket{ab}-\eta\ket{ba})(\bra{ab}-\eta\bra{ba})$.
$h'_j$ is a generic Hermitian operator with support on the vicinity of the site $j$. 

We highlight an important feature of the XXC model, which is not only closely related to its integrability, but also plays a crucial role in hosting the generalized spin helix states. We notice that the local Hamiltonian $h_{j,j+1}$ in Eq.~\eqref{eq:Multi} can be deformed into a \textit{non-Hermitian} operator:
\begin{align}
\tilde{h}_{j,j+1}&=h_{j,j+1}+i\sin(\gamma)\sum_{a\in A} (E_{aa}^j\otimes I_{j+1}-I_j\otimes E_{aa}^{j+1})\nonumber\\
&=\sum_{a\in A,b\in B} [\eta_{b} E^j_{ab}\otimes E^{j+1}_{ba}+\eta_{b}^{-1} E^j_{ba}\otimes E^{j+1}_{ab}\nonumber\\
& \qquad -e^{-i\gamma}E^j_{aa}\otimes E^{j+1}_{bb} -e^{i\gamma} E^j_{bb}\otimes E^{j+1}_{aa}],
\end{align}
where $I_j$ denotes the identity operator on site $j$. For the one-dimensional chain with the periodic boundary condition, the summation of $\tilde{h}_{j,j+1}$ yields the same Hamiltonian $H_{\text{XXC}}$ due to the consecutive cancellation of the on-site non-Hermitian terms. We observe that the operators $e_j=\tilde{h}_{j,j+1}$ satisfy the (periodic) Temperley-Lieb algebra relations \cite{Temperley1971Relations,martin1994blob}: $e_j^2=-2\cos(\gamma)e_j$, $e_je_{j\pm 1}e_j=e_j$, and $e_je_{j'}=e_{j'}e_j$ for $|j-j'|>1$. The second relation is a reformulation of the Yang-Baxter equations \cite{Baxter2016Exactly}. The first condition implies that $\tilde{h}_{j,j+1}$ is proportional to a non-Hermitian projector. The local null space of $\tilde{h}_{j,j+1}$ is larger than that of $h_{j,j+1}$: apart from the simple states $\ket{a,a}_{j,j+1}$ and $\ket{b,b}_{j,j+1}$, there exist one more state $\frac{1}{\sqrt{2}}[\ket{a,b}-\eta_{b}\exp(-i\gamma)\ket{b,a}]_{j,j+1}$ that are also annihilated by $\tilde{h}_{j,j+1}$, for any $a\in A,b\in B$. 

For the non-integrable Hamiltonian Eq.~\eqref{eq:perturbed}, we can identify several Krylov subspaces that are preserved under the action of the Hamiltonian. A Krylov subspace labeled by $\mathcal{H}_{a,b}$ is spanned by computational basis states with only two kinds of local bases $\ket{a}$ and $\ket{b}$, for $a\in A$ and $b\in B$. Projected into the subspace $\mathcal{H}_{a,b}$, the Hamiltonian Eq.~\eqref{eq:perturbed} is reduced to the spin-$1/2$ XXZ Hamiltonian between $\ket{a}$ and $\ket{b}$, thus making the subspace integrable. The complementary subspace of all the integrable subspaces is non-integrable due to the presence of $Ph'P$ terms. Therefore, the many-body Hilbert space fragments into several integrable subspaces and one non-integrable subspace.

By combining the non-Hermitian Hamiltonian densities $\tilde{h}_{j,j+1}$ and Hermitian projectors $P_{j,j+1}$, we find the following generalized spin helix states in the $E=0$ subspace of the perturbed Hamiltonian $H$ as exact quantum many-body scars:
\begin{align}\label{eq:generalizedSHS}
    &\ket{\Psi(\{\beta_c\}_{c=1}^N)}=\bigotimes_{j=1}^L 
\ket{\psi_j},\nonumber\\
    &\ket{\psi_j}=\sum_{a\in A}(\beta_a\ket{a}_j)+e^{i(j-1)\gamma}\sum_{b\in B} (\eta_b^{j-1}\beta_b\ket{b}_j).
\end{align}
Here, $\{\beta_c\}_{c=1}^N$ is a set of arbitrary complex parameters. Due to the periodic boundary condition, we require $\gamma L/2\pi$ to be an integer. 

Several remarks come in order. First, the generalized spin helix states span the scar subspace overcompletely. To extract complete orthogonal bases for the scar subspace, we expand the product state as a multi-variable polynomial with respect to the parameters $\{\beta_c\}_{c=1}^N$:
\begin{equation}
    \ket{\Psi(\{\beta_c\}_{c=1}^N)}=\sum_{\begin{subarray}{c}\{m_c\}_{c=1}^N,\\ \sum_c m_c=L \end{subarray}}(\prod_{c=1}^N \beta_c^{m_c})\ket{S_{m_1,\cdots,m_N}},
\end{equation}
such that the ``coefficients" of the polynomial are orthogonal scar states $\ket{S_{m_1,\cdots,m_N}}$ labeled by the powers in the associated terms $\prod_{c=1}^N \beta_c^{m_c}$. The procedure of unfolding scar states according to certain parameters can be regarded as the reversal of the compressed MPS technique used to find local annihilators for a tower of scar states \cite{Shibata2020Onsager,Chattopadhyay2020quantum,Mark2020Unified,Wang2023Embedding}. 
The total dimension of the scar subspace can be determined through combinatorially counting all the possible non-negative integer arrays $\{m_c\}_{c=1}^N$ that sum up to $L$, resulting in the binomial coefficient $\binom{L+N-1}{N-1}$. For large $L$, the dimension scales as $L^{N-1}$, which is significantly smaller than the dimension of the entire Hilbert space $N^L$, fulfilling the definition of weak violation of ETH.

Second, we emphasize that our construction of the generalized spin helix states as zero-energy scars is distinct from the Shiraishi-Mori embedding method~\cite{Shiraishi2017Systematic}, since $[P_{j,j+1},H_{\text{XXC}}]\neq 0$ in Eq.~\eqref{eq:perturbed}. The key aspect of our construction lies in the crucial observation that $H_{\text{XXC}}$ can be expressed as the sum of underlying local non-Hermitian projectors. In previous studies, the Hamiltonians hosting a tower of exact scars typically include two ingredients: 1) Hamiltonian densities locally annihilating all the scar states, which can be obtained by the compressed MPS technique; 2) Hamiltonians that lift the degeneracy of the scar states according to their different values of a good quantum number~\cite{Mark2020Unified,Moudgalya2020eta}. However, it has been discovered that certain scarred Hamiltonians annihilate the scar states as a whole rather than locally \cite{Mark2020Unified,Mark2020eta,nicholas2020from, Omiya2023Fractionalization}, which are referred to as ``type II symmetric Hamiltonians" in Ref.~\cite{Moudgalya2022Exhaustive}.
The integrable XXC Hamiltonian annihilates the generalized spin helix states as a whole and exactly 
serves as an example of this kind of Hamiltonians. In \cite{Supp} we show that most of these ``annihilate-as-a-whole" Hamiltonians (e.g., the spin-$1$ Affleck-Kennedy-Lieb-Tasaki model) can be unified into the same framework by decomposing them into the summation of \textit{non-Hermitian} densities, which annihilate the scar tower states \textit{locally}. 
We further propose a systematic method to construct such special Hamiltonians based on the scar tower states, which highlights the necessity of introducing the non-Hermitian decomposition method. In the following we specifically discuss several models that can be unified into our constructions.



\begin{figure}
\centering
\includegraphics[width=0.98\linewidth]{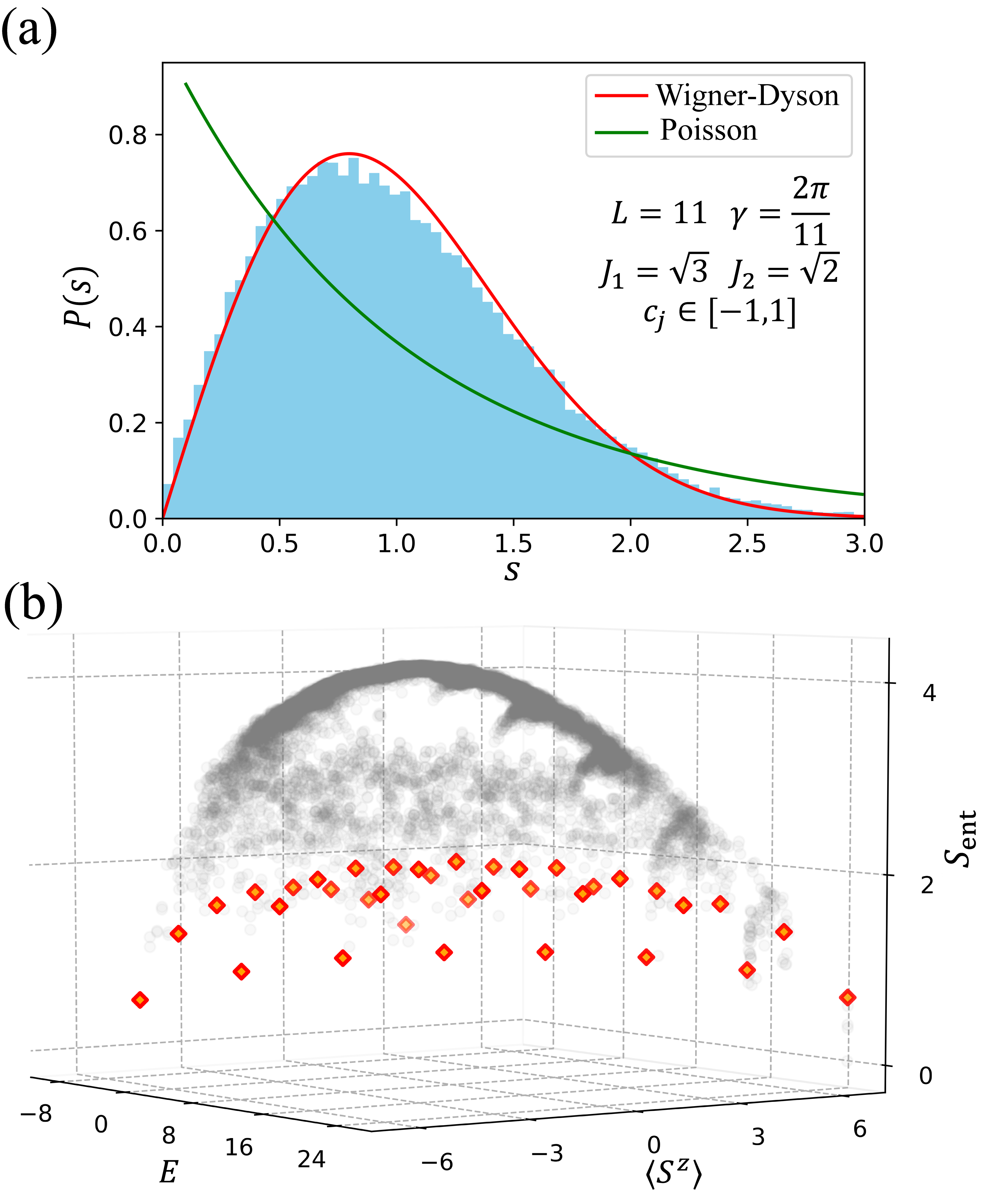} 
\caption{(a) Level spacing statistics of the non-integrable subspace of the $N=3$ partially integrable model.  (b) Bipartite entanglement entropy of the thermal  eigenstates (gray dots) and scarred eigenstates (red diamonds) within the non-integrable subspace, as a function of the eigenenergy and the expectation value of total $S^z$. $L=9$, $\gamma=2\pi/9$, other parameters taken as the same as those in (a).
}
\label{fig:Level_stat_Entropy}
\end{figure}

\textit{The $N=3$ case.}-- Consider the case of $N=3, N_A=1, N_B=2$ without twisting ($\eta_{b}=1$). We take the partition $A=\{1\}, B=\{2,3\}$ and map the model to the spin-$1$ chain through the correspondence $\ket{c=1}\to\ket{0},\ket{c=2}\to\ket{1}$ and $\ket{c=3}\to\ket{-1}$. We represent the non-Hermitian term added to $h_{j,j+1}$ as $i\sin(\gamma)[-(S_j^z)^2+(S_{j+1}^z)^2]$, where $S^\alpha_j\  (\alpha=x,y,z)$ denotes the spin-$1$ operator on the site $j$. The corresponding Hamiltonian Eq.~\eqref{eq:perturbed} hosts the zero-energy generalized spin helix states 
$\ket{\Psi(\beta_+,\beta_-)}=\bigotimes_{j=1}^L \{\beta_+\exp[i(j-1)\gamma]\ket{1}_j + \ket{0}_j + \beta_-\exp[i(j-1)\gamma]\ket{-1}_j\}$.
The unfolded scar states $\ket{S_{m_1,m_2,m_3}}$ are labeled by two independent integers (since $m_1+m_2+m_3=L$). Scarred eigenstates with $m_2=0$ or $m_3=0$ belong to the integrable subspaces $\mathcal{H}_{0,\pm}$ and are actually the same as the spin-$1/2$ helix states. On the other hand, when $m_1=0$, the resulting scar states are built by creating $m_2$ bi-magnons (flipping $\ket{-1}$ to $\ket{1}$) with zero momentum on the reference state $\ket{S_{0,0,L}} = \otimes_{j=1}^L \ket{-1}_j $ (i.e., Dicke states for $\ket{\pm 1}$ ). In addition to these special cases, there exist many scarred eigenstates simultaneously possessing the three basis states $\{\ket{-1}, \ket{0}, \ket{1} \}$ within the non-integrable subspace. 
In order to break the energy degeneracy of the scar subspace, we add the terms $\sum_j [J_1 S_j^z + J_2 (S_{j}^z)^2]$, leading to the energy splitting as $E_{m_1,m_2,m_3}=(J_1+J_2)m_2 - (J_1-J_2)m_3$. As shown in Fig.~\ref{fig:Level_stat_Entropy}(a), we specifically take $h'_{j} = c_j S^x_{j-1}$ ($c_j$ is a site-dependent random coefficient) for the integrability breaking $Ph'P$ terms and compute the statistics of the energy level spacings $s$ in the non-integrable subspace. The distribution $P(s)$ follows the Wigner-Dyson distribution well with the level-spacing ratio $\langle r \rangle = 0.5245$, which confirms its chaotic nature~\cite{Pal2010Manybody}. Fig.~\ref{fig:Level_stat_Entropy}(b) displays the bipartite entanglement
entropy $S_\text{ent}$ of all the eigenstates within the non-integrable subspace, highlighting the atypical scarred eigenstates with red diamonds.
In \cite{Supp} we have also obtained the analytical expression of $S_\text{ent}$ for the scar states $\ket{S_{m_1,m_2,m_3}}$, and found that it at most follows the sub-volume-law scaling $S_\text{ent} \sim \ln L$.



\textit{Spinful fermion representation for $N=4$.}-- Our constructions of scarred models hosting generalized spin helix states can be extended to fermionic degrees of freedom. Specifically, we consider following $SU(2)$-symmetric fermionic model: 
\begin{equation}\label{eq:electronic}
    h_{j,j+1}=\sum_{\sigma=\uparrow,\downarrow}(c_{j,\sigma}^\dagger c_{j+1,\sigma}+c_{j+1,\sigma}^\dagger c_{j,\sigma})-\cos(\gamma)V(n_j,n_{j+1}),
\end{equation}
where $c_{j,\sigma}^\dagger (c_{j,\sigma})$ is the fermionic creation (annihilation) operator with spin $\sigma$ at site $j$. $n_j = n_{j,\uparrow} + n_{j,\downarrow}$, $n_{j,\sigma} = c_{j,\sigma}^\dagger c_{j,\sigma} $. The nearest-neighbor interaction is given by $V(n_j,n_{j+1})=(n_j-n_{j+1})^2(2-n_j-n_{j+1})^2$, which could be realized by engineering Coulomb repulsion between ultracold spinful fermions on optical lattices \cite{Dutta2015Non}.

To demonstrate how Eq.~\eqref{eq:electronic} fits in our scarred models Eq.~\eqref{eq:perturbed} as the $N=4, N_A=N_B=2$ case, first we establish the correspondence between the bosonic basis $\ket{c}$ and the fermionic basis as follows: $\ket{c=1}\to \ket{0}, \ket{c=2}\to \ket{\uparrow}=c_{\uparrow}^\dagger\ket{0}, \ket{c=3}\to\ket{\downarrow}=c_{\downarrow}^\dagger\ket{0}, \ket{c=4}\to\ket{d}=c_{\uparrow}^\dagger c_{\downarrow}^\dagger\ket{0}$, and $A=\{2,3\},B=\{1,4\}$. Here $\ket{0}$ is the empty state and $\ket{d}$ is the doubly occupied state. 
Notice that total fermion number on the two sites $n=n_j+n_{j+1}$ is a conserved quantity of $h_{j,j+1}$. Therefore, the local Hamiltonian can be decomposed to blocks with definite $n$, denoted as $h_{j,j+1}^n$. For $n=1$, two pairs of basis states $[\ket{\uparrow},\ket{0}]_{j,j+1}$ and $[\ket{\downarrow},\ket{0}]_{j,j+1}$ are involved. In this case, the single-fermion hopping and interaction terms exactly match the corresponding terms in the XXC model, without twisting ($\eta_0=1$). Similarly, the $n=3$ block consists of pairs $[\ket{\uparrow},\ket{d}]_{j,j+1}$ and $[\ket{\downarrow},\ket{d}]_{j,j+1}$, but with a twisting factor $\eta_d=-1$ originating from the fermion anti-commutation relation. 
Hamiltonian densities of the above two blocks sum to the integrable $N=4$ XXC model \cite{Dargis1998Fermionization}, 
and their non-Hermitian deformation 
can be constructed as $i\sin(\gamma)(\ket{\uparrow}\bra{\uparrow}_j+\ket{\downarrow}\bra{\downarrow}_j-\ket{\uparrow}\bra{\uparrow}_{j+1}-\ket{\downarrow}\bra{\downarrow}_{j+1})=i\sin(\gamma)(n_j-n_{j+1})(2-n_j-n_{j+1})$.

The remaining non-zero block of Eq. \eqref{eq:electronic} is that with $n=2$. Intriguingly, $h_{j,j+1}^{n=2}$ itself breaks the integrability and serves as the perturbation term $P_{j,j+1}h'_jP_{j,j+1}$. Note that the local null space of $h_{j,j+1}^{n=2}$ includes states $(\ket{\uparrow,\downarrow}+\ket{\downarrow,\uparrow})_{j,j+1}$ and $(\ket{0,d}-\ket{d,0})_{j,j+1}$, which makes $h_{j,j+1}^{n=2}$ locally annihilate the following generalized spin helix states:
\begin{align}
    \ket{\Psi (\beta_\uparrow, \beta_\downarrow, \beta_d)} & =  \bigotimes_{j=1}^L ( \ket{0}_j + \beta_\uparrow e^{i(j-1)\gamma} \ket{\uparrow}_j \nonumber \\
    &+ \beta_\downarrow e^{i(j-1)\gamma} \ket{\downarrow}_j + \beta_d (-1)^{j-1} \ket{d}_j ).
\label{eq:ElectronicHelix}
\end{align}
To sum up, we have proved that the spinful fermionic Hamiltonian Eq.~\eqref{eq:electronic} is a realization of the $N=4$ partially integrable model Eq.~\eqref{eq:perturbed} hosting the unfolded orthogonal scar states $\ket{ S_{m_1,m_2,m_3,m_4} }$. Remarkably, the states with $m_2=m_3=0$ (in the integrable subspace $\mathcal{H}_{0,d}$) are known as the $\eta$-pairing states of the Fermi-Hubbard model \cite{Yang1989Pairing}, which have been realized as scar states in several non-integrable models \cite{Mark2020eta,Moudgalya2020eta}. Adding chemical potential, magnetic field pointing to the spin-$z$ direction, and Hubbard interaction can completely lift the degeneracy of scar subspace.


\textit{Higher-spin helix states in the clock models.}-- The above constructions of partially integrable scarred Hamiltonians are all based on the spin-$1/2$ XXZ interaction and the corresponding helix states. It is also possible to exploit higher-spin interactions and their helix states as the building blocks. 
We now shift our focus to the $U(1)$-invariant clock models proposed in Ref. \cite{Vernier2019Onsager}. The local Hamiltonian is given by (up to a constant)
\begin{align*}
    h_{j,j+1}=&\sum_{a=1}^{M-1} \frac{1}{2\sin(a\gamma)}[M(-1)^a (S_j^{-}S_{j+1}^+)^a+\text{H.c.}\nonumber\\
    &+(M/2-a)\exp(ia\gamma)(\tau_j^a+\tau_{j+1}^a-2)],
\end{align*}
where $\gamma=\pi/M$. $S_j^\pm$ and $\tau_j$ act on the local Hilbert space as $ S_j^+\ket{a}_j=(1-\delta_{a+1,M})\ket{a+1}_j, \tau_j\ket{a}_j=\exp(2ia\gamma)\ket{a}_j$.
The total Hamiltonian commutes with the $U(1)$ charge operator $Q=\sum_{j=1}^L Q_j=\sum_{j=1}^{L} \sum_{a=1}^{M-1}\exp(ia\gamma)\tau_j^a/2i\sin(a\gamma)$.

We identify the spin helix states in this model by performing the non-Hermitian deformation as $\tilde{h}_{j,j+1}=h_{j,j+1}+iM(Q_j-Q_{j+1})/2$. By the similar method in finding the local null states of the spinful-fermion model Eq.~\eqref{eq:electronic}, we decompose the local non-Hermitian Hamiltonian $\tilde{h}_j$ into blocks with definite $U(1)$ charge $q=Q_j+Q_{j+1}$. In each block we discover a two-local null state as $\sum_{p=0}^q \exp(-ip(\pi-\gamma))\ket{p,q-p}_{j,j+1}$, which constitutes the following higher-spin helix states 
\begin{equation}
\ket{\Psi(\beta)}=\bigotimes_{j=1}^L(\sum_{p=0}^{M-1}\beta^p e^{ip(j-1)(\pi-\gamma)} \ket{p}_j).
\end{equation}

By utilizing the clock models and their higher-spin helix states as a starting point, we can construct more similar partially integrable scarred Hamiltonians. As a minimal example, we can consider embedding two sets of $M=3$ spin helix states into a model with four local degrees of freedom, labeled by $\ket{0},\ket{\bar{0}}$ and $\ket{1},\ket{2}$. Here, $\ket{0}(\ket{\bar{0}})$ interacts with  $\ket{1}$ and $\ket{2}$ as the $M=3$ clock model. We can further add the $Ph'P$ perturbation terms between the basis states $\ket{0}$ and $\ket{\bar{0}}$ on neighboring sites like Eq.~\eqref{eq:projector} to break the integrability, while preserving the generalized spin helix states as quantum many-body scars.

\textit{Conclusions.}-- 
In summary, we have presented a general method to construct a variety of partially integrable Hamiltonians hosting the generalized spin helix states as exact quantum many-body scars.
Our work further demonstrates that the philosophy of adding designed integrability-breaking terms to exactly solvable models serves as a promising avenue for systematic constructions of scarred Hamiltonians. 
Moreover, inspired by the Temperley-Lieb algebra relations satisfied by the non-Hermitian deformation of the XXC Hamiltonian, we decompose several extensive local scarred Hamiltonians into the summation of local non-Hermitian annihilators.
This method gives a unified framework to understand these special types of ``annihilate-as-a-whole" scarred Hamiltonians, which could have intriguing connections with the language of commutant algebra~\cite{Moudgalya2022Exhaustive,Moudgalya2022Hilbert}.

We acknowledge helpful discussions with Berislav Bu\v{c}a, Zhong Wang, Li-Wei Yu, Shun-Yao Zhang, Dong-Ling Deng, and Fei Song. 

\bibliography{Heran_Helix,QMBS,DengQAIGroup}
\end{document}


\title{Supplementary Materials for: Generalized Spin Helix States as Quantum Many-Body Scars in Partially Integrable Models}

\author{He-Ran Wang}
\email{whr21@mails.tsinghua.edu.cn}
\affiliation{Institute for Advanced Study, Tsinghua University, Beijing 100084, People's Republic of China}

\author{Dong Yuan}
\email{yuand21@mails.tsinghua.edu.cn}
\affiliation{Center for Quantum Information, IIIS, Tsinghua University, Beijing 100084, People's Republic of China}

\maketitle

\section{Temperley-Lieb relations of the XXC Hamiltonian}
In this section, we prove that the non-Hermitian deformation of the local XXC Hamiltonian follows the Temperley-Lieb relations. That is, the operator $e_j$ defined as
\begin{equation}
    e_j=\sum_{a\in A,b\in B}(\eta_{b} E^j_{ab}\otimes E^{j+1}_{ba}+\eta_{b}^{-1} E^j_{ba}\otimes E^{j+1}_{ab}-\exp(-i\gamma)E^j_{aa}\otimes E^{j+1}_{bb}-\exp(i\gamma)E^j_{bb}\otimes E^{j+1}_{aa})=\sum_{a\in A,b\in B}e_j^{a,b}
\end{equation}
satisfies the relations $e_j^2=-2\cos(\gamma)e_j, e_je_{j\pm 1}e_j=e_j$ and $e_je_{j'}=e_{j'}e_j, \forall |j-j'|>1$. To prove the first relation, notice that every single term $e_j^{a,b}$ is orthogonal mutually, and $(e_j^{a,b})^2=-2\cos(\gamma)e_j^{a,b}$. For the third one, the proof is straightforward since the support of $e_j$ only covers the site $j$ and $j+1$. We show the lengthy derivations about the second relation:
\begin{align}
e_je_{j+1}=&\sum_{a,a'\in A;b,b'\in B}e_j^{a,b}e_{j+1}^{a',b'}\nonumber\\
=&\sum_{a,a'\in A;b\in B}(\eta_b^{-2}E^j_{ba}\otimes E^{j+1}_{aa'}\otimes E_{a'b}^{j+2}+E_{aa}^j\otimes E_{bb}^{j+1}\otimes E_{a'a'}^{j+2}\nonumber\\
&-\exp(i\gamma)\eta_b^{-1} E^j_{ba}\otimes E^{j+1}_{ab} \otimes E_{a'a'}^{j+2}-\exp(-i\gamma)\eta_b^{-1} E_{aa}^j\otimes E_{ba'}^{j+1}\otimes E_{a'b}^{j+2})\nonumber\\
&+\sum_{a\in A;b,b'\in B}(\eta_b\eta_{b'}E^j_{ab}\otimes E^{j+1}_{bb'}\otimes E_{b'a}^{j+2}+E_{bb}^j\otimes E_{aa}^j\otimes  E_{b'b'}^{j+2}\nonumber\\
&-\exp(-i\gamma)\eta_b E^j_{ab}\otimes E^{j+1}_{ba} \otimes E_{b'b'}^{j+2}-\exp(i\gamma)\eta_{b'}E_{bb}^j\otimes E_{ab'}^{j+1}\otimes E_{b'a}^{j+2}),
\end{align}
\begin{align}
e_je_{j+1}e_j=&\sum_{a,a'\in A;b\in B}(\eta_b E_{ab}^j\otimes E_{ba}^{j+1}\otimes E_{a'a'}^{j+2}-\exp(-i\gamma)E_{aa}^j\otimes E_{bb}^{j+1}\otimes E_{a'a'}^{j+2}\nonumber\\
&-\exp(i\gamma) E^j_{bb}\otimes E^{j+1}_{aa} \otimes E_{a'a'}^{j+2}+\eta_b^{-1} E_{ba}^j\otimes E_{ab}^{j+1}\otimes E_{a'a'}^{j+2})\nonumber\\
&+\sum_{a\in A;b,b'\in B}(\eta_b^{-1}E_{ba}^j\otimes E_{aa}^j\otimes  E_{b'b'}^{j+2}-\exp(i\gamma)E_{bb}^j\otimes E_{aa}^j\otimes  E_{b'b'}^{j+2}\nonumber\\
&-\exp(-i\gamma) E^j_{aa}\otimes E^{j+1}_{bb} \otimes E_{b'b'}^{j+2}+\eta_b E_{ab}^j\otimes E_{ba}^{j+1}\otimes E_{b'b'}^{j+2})\nonumber\\
=&(\sum_{a\in A,b\in B}e_j^{ab})\otimes(\sum_{a\in A}E_{aa}^{j+2}+\sum_{b\in A}E_{bb}^{j+2})=e_j.
\end{align}

\section{Extensive local annihilators of typical scarred models}
In this section, we provide more discussions about the extensive local Hamiltonians annihilating a tower of given scar states.
By ``extensive local'', we refer to the operators as summations of translational-invariant strictly local operators with finite support.
We mainly focus on those extensive local annihilators that cannot be decomposed to local Hermitian annihilators, and thus only act as a whole.
Ref. \cite{Mark2020Unified} noticed that the spin-1 AKLT Hamiltonian (minus the total spin-$z$ Hamiltonian) is such an extensive local annihilator of a tower of scars (see also Sec. \ref{sec:AKLT}). 
Subsequently, the similar Hamiltonians in the spin-$1$ $XY$ model and the generalized Fermi-Hubbard model have been proposed in Ref. \cite{Mark2020eta}, where the authors stated that those extensive local Hamiltonians are exceptions to the Shiraishi-Mori embedding (Sec. \ref{sec:XY} and Sec. \ref{sec:eta}). Ref. \cite{nicholas2020from} presented a numerical method (the generalized covariance-matrix algorithm) to search the extensive local annihilators for several models.

Two unified frameworks have been proposed for understanding the extensive local annihilators.
First, Moudgalya and Motrunich established the mathematical relation between scar states and scarred Hamiltonians as mutually commutant algebras, where the extensive local annihilators are dubbed as ``type II symmetric operators" and can only be constructed case-by-case \cite{Moudgalya2022Exhaustive}.
The other is Ref. \cite{Omiya2023Fractionalization}, where the authors enlarged the local Hilbert space, then the extensive local annihilators in the enlarged Hilbert space fit into the local embedding formalism. 
Different from aforementioned two frameworks, we decompose the extensive local annihilators into local non-Hermitian projectors or annihilators, e.g., the XXC Hamiltonian in the main text. 
In the following Secs. \ref{sec:DMI}-\ref{sec:AKLT}, we review several known scarred eigenstates and their extensive local annihilators, and explicitly show how the non-Hermitian decomposition applies to those models. Finally, in Sec. \ref{sec:criterion}, we propose a systematic method to find such extensive local annihilators for a given tower of scar states.


\subsection{Spin-$1/2$ Dzyaloshinskii-Moriya Interaction}\label{sec:DMI}
We begin with the Dzyaloshinskii-Moriya Interaction (DMI) terms defined on a spin-$1/2$ chain under PBC: $H_{\alpha-\text{DMI}}=\sum_{j=1}^L (\vec{S}_j\times \vec{S}_{j+1})\cdot\hat{\alpha}$. The DMI Hamiltonian as a whole annihilates the ferromagnetic Dicke states \cite{Mark2020eta,Dooley2021Robust,Tang2022Multimagnon,Moudgalya2022Exhaustive}. These states can be embedded as the scarred eigenstates in the toy model introduced by Ref. \cite{Choi2019emergent}, and will persist as scars after adding the DMI terms. Two-local components of these Dicke scar states are spin-$1$ triplets formed by two neighboring spin-$1/2$'s, so the toy model in~\cite{Choi2019emergent} is constructed through the Shiraishi-Mori embedding method~\cite{Shiraishi2017Systematic} by using the local projectors onto the spin singlets. We show that the local DMI term does the same projection after the following non-Hermitian deformation.

Without loss of generality, we set $\hat{\alpha}=\hat{z}$, such that the local DMI term is given by
\begin{equation}\label{eq:DMI}
    h_{j,j+1}=S_j^xS_{j+1}^y-S_j^yS_{j+1}^x=\frac{i}{2}(S_j^{+}S_{j+1}^{-}-S_j^{-}S_{j+1}^{+})=\frac{i}{2}(\ket{\uparrow\downarrow}\bra{\downarrow\uparrow}-\ket{\downarrow\uparrow}\bra{\uparrow\downarrow})_{j,j+1}.
\end{equation}
We choose the non-Hermitian deformation as 
\begin{equation}\label{eq:DMI}
    \tilde{h}_{j,j+1}=h_{j,j+1}-\frac{i}{2}(S_j^z-S_{j+1}^z)=\frac{i}{2}[(\ket{\uparrow\downarrow}+\ket{\downarrow\uparrow})(\bra{\downarrow\uparrow}-\bra{\uparrow\downarrow})]_{j,j+1}.
\end{equation}
The non-Hermitian operator $\tilde{h}_{j,j+1}$ transforms the spin singlet to the triplet with zero total spin-$z$ component. Evidently, $\tilde{h}_{j,j+1}$ annihilates the Dicke scar states locally, and they sum to the Hermitian DMI Hamiltonian under PBC. Another equivalent viewpoint of the annihilation is to regard the Dicke states as the condensation of zero-momentum magnons, which undergo destructive interference through the DMI terms. In the following, we will see that other extensive local Hamiltonians hosting exact scar tower states resemble the similar structure of DMI terms.

\subsection{Twisted DMI}
Considering twisting the spin-$1/2$ DMI Hamiltonian through the unitary gauge transformation $U=\otimes_j e^{-ij\gamma S_j^z}$, which leads to
\begin{equation}
    U h_{j,j+1}U^\dagger=\frac{i}{2}(e^{i\gamma}S_j^{+}S_{j+1}^{-}-e^{-i\gamma}S_j^{-}S_{j+1}^{+})=\frac{i}{2}(e^{i\gamma}\ket{\uparrow\downarrow}\bra{\downarrow\uparrow}-e^{-i\gamma}\ket{\downarrow\uparrow}\bra{\uparrow\downarrow})_{j,j+1}.
\end{equation}
And accordingly, the same non-Hermitian deformation can be applied as
\begin{equation}
    U\tilde{h}_{j,j+1}U^\dagger=U h_{j,j+1}U^\dagger-\frac{i}{2}(S_j^z-S_{j+1}^z)=\frac{i}{2}[(e^{i\gamma/2}\ket{\uparrow\downarrow}+e^{-i\gamma/2}\ket{\downarrow\uparrow})(e^{i\gamma/2}\bra{\downarrow\uparrow}-e^{-i\gamma/2}\bra{\uparrow\downarrow})]_{j,j+1}.
\end{equation}
Notably, $U\tilde{h}_{j,j+1}U^\dagger$ locally annihilates the spin helix states of spin-$1/2$ XXZ model. Therefore, both the twisted DMI Hamiltonian $U_\gamma H_{z-\text{DMI}}U_\gamma^\dagger$ and the XXZ Hamiltonian serve as extensive local annihilators as the spin helix states \cite{Shi2023Robust}, though the non-Hermitian deformations of two models are different.

\subsection{Spin-$1$ $XY$ model}\label{sec:XY}
Next, we consider the spin-$1$ $XY$ model $H_{XY} = \sum_{j} [ S^x_j S^x_{j+1} + S^y_j S^y_{j+1} + h S^z_j + D (S^z_j)^2 ]$ \cite{Schecter2019Weak}, with $L+1$ scarred eigenstates generated from the ferromagnetic state $|S_0\rangle=\ket{-1,\cdots,-1}$ by acting the ladder operator $Q^\dagger=\sum_j (-1)^j(S_j^{+})^2$. Ref. \cite{Mark2020eta} discussed an extensive local Hamiltonian annihilating the scar states:
\begin{equation}\label{Spin1XY}
    H=\sum_j h_{j,j+1}=\sum_j i(\ket{-1,1}\bra{1,-1}-\ket{1,-1}\bra{-1,1})_{j,j+1}=\sum_j \frac{i}{4}[(S_j^-)^2(S_{j+1}^+)^2-(S_{j}^+)^2(S_{j+1}^-)^2].
\end{equation}
Note that $h_{j,j+1}$ in Eq. \eqref{Spin1XY} is indeed the effective DMI term between $\ket{1}$ and $\ket{-1}$. Subsequently, the non-Hermitian deformation can be constructed as
\begin{equation}
    \tilde{h}_{j,j+1}=h_{j,j+1}+\frac{i}{2}[-S_j^z-(S_j^z)^2+S_{j+1}^z+(S_{j+1}^z)^2]=i[(\ket{-1,1}-\ket{1,-1})(\bra{-1,1}+\bra{1,-1})-\ket{1,0}\bra{1,0}+\ket{0,1}\bra{0,1}]_{j,j+1}.
\end{equation}
Here $\tilde{h}_{j,j+1}$ is the summation of two orthogonal projectors. Specifically, $\tilde{h}_{j,j+1}$ annihilates $(\ket{-1,1}-\ket{1,-1})_{j,j+1}$, thus annihilating the whole scar tower of the spin-1 $XY$ model. 

\subsection{$\eta$-pairing states} \label{sec:eta}
In this subsection we will investigate the $\eta$-pairing states, which are closely related to the $\eta$-pairing symmetry of the Hubbard model on the bipartite lattice \cite{Yang1989Pairing}. They are generated by acting $Q^\dagger=\sum_j (-1)^j c_{j,\uparrow}^\dagger c_{j,\downarrow}^\dagger$ on the electron vacuum $|\Omega\rangle$ repeatedly. The same scar states are also discovered in our $N=4$ model in the main text. Recently, some other Hamiltonians hosting the $\eta$-pairing states while breaking the pairing symmetry are searched and constructed \cite{Mark2020eta,Pakrouski2021Group,Kolb2023Stability}. Among them, we are interested in the following extensive local Hamiltonian 
\begin{equation}\label{eq:fermionDMI}
    H=i\sum_j (c_{j+1,\uparrow}^\dagger c_{j+1,\downarrow}^\dagger c_{j,\downarrow}c_{j,\uparrow}- c_{j,\uparrow}^\dagger c_{j,\downarrow}^\dagger c_{j+1,\downarrow}c_{j+1,\uparrow}).
\end{equation}
Locally, it sends $\ket{d,0}$ to $\ket{0,d}$ and vice versa, with opposite amplitude. This form of the local Hamiltonian implies the DMI-type mechanism. We aim to annihilate the local state $(\ket{0,d}-\ket{d,0})_{j,j+1}$, so the deformation is given by 
\begin{equation}
    \tilde{h}_{j,j+1}=h_{j,j+1}+i(n_{j+1,\uparrow}n_{j+1,\downarrow}-n_{j,\uparrow}n_{j,\downarrow}).
\end{equation}

We can show the relations between the fermionic Hamiltonian Eq. \eqref{eq:fermionDMI} and DMI Hamiltonian more clearly by applying the particle-hole transformation, sending $c_{j,\downarrow}\to (-1)^jc_{j,\downarrow},c_{j,\downarrow}^\dagger\to (-1)^jc_{j,\downarrow}^\dagger $. Then we define the local spin-1/2 operator as the bilinear form of fermions as $S_j^+=c_{j,\uparrow}^\dagger c_{j,\downarrow},S_j^-=c_{j,\downarrow}^\dagger c_{j,\uparrow}$, the resulting extensive local Hamiltonian is given by
\begin{equation}
     H=i\sum_j(S_j^{+}S_{j+1}^{-}-S_j^{-}S_{j+1}^{+}),
\end{equation}
which shares the same form as Eq. \eqref{eq:DMI}. Accordingly, the $\eta$-pairing states are transformed to the Dicke states with one fermion occupied per site.

\subsection{Spin-1 AKLT model} \label{sec:AKLT}
In this subsection, we consider the spin-1 Affleck-Kennedy-Lieb-Tasaki (AKLT) model $H_{\text{AKLT}}=\sum_j T_{j,j+1}^{S=2}$, where $T_{j,j+1}^{S=2}$ projects two adjacent spin-$1$'s onto a total spin-$2$~\cite{Affleck1987Rigorous}. A tower of scarred eigenstates is generated from the ground state $|S_0\rangle=|G\rangle$ by the ladder operator $Q^\dagger=\sum_j (-1)^j(S_j^{+})^2 $, $\ket{S_n}=(Q^\dagger)^n\ket{S_0}$~\cite{Moudgalya2018Exact,Moudgalya2018Entanglement}. The parent state $\ket{S_0}$ admits the MPS representation:
\begin{equation}\label{eq:AKLTGS}
\ket{S_0}=\sum_{\mu_1,\mu_2,\cdots,\mu_L}\textnormal{Tr}\left[A^{(\mu_1)}A^{(\mu_2)}\cdots A^{(\mu_L)}\right]\ket{\mu_1,\mu_2,\cdots,\mu_L},
\end{equation}
where $\mu=\pm 1$ or $0$, $A^{(\pm 1)}=\mp\sqrt{\frac{2}{3}}\sigma^\mp$ and $A^{(0)}=-\frac{1}{\sqrt{3}}\sigma^z$. For later convenience, we also introduce the notation of $k$-local MPS as
\begin{equation}
    \ket{M(l,r)}=\sum_{\mu_1,\cdots,\mu_k}(A^{(\mu_1)} \cdots A^{(\mu_k)} )_{l,r} \ket{\mu_1,\cdots,\mu_k},
\end{equation}
where $l,r$ are the left and right uncontracted indices (dangling bonds) of the local MPS.  

Before dividing into detailed calculations, we clarify different concepts of type-II symmetric Hamiltonians in the AKLT model \cite{Moudgalya2022Exhaustive}. First, the ground state $\ket{S_0}$ is annihilated by the total spin-$z$ operator $S^z_{\text{tot}}=\sum_j^L S_j^z$, which cannot be decomposed to local Hermitian annihilators. In that sense, $S^z_{\text{tot}}$ is a type-II Hamiltonian for the AKLT ground state. On the other hand, we can consider the extensive local annihilators of the whole scar tower $\{\ket{S_n}\}_{n=0}^{L/2}$: It can be shown that $H_{\text{AKLT}}\ket{S_n}=S^z_{\text{tot}}\ket{S_n}=2n\ket{S_n}$ , therefore $H_{\text{AKLT}}-S^z_{\text{tot}}$ is a type-II Hamiltonian for the AKLT scar tower. In the following, we will discuss the non-Hermitian deformation for the both cases.

For the ground state, we explicitly find that $S_j^z\ket{S_0}=(O_{j-1,j}-O_{j,j+1})\ket{S_0}$, where $O_{j,k}$ is a two-local non-Hermitian operator. The rigorous analysis goes as follows: The action of $S_j^z$ on $\ket{S_0}$ manifests as a single-mode excitation in the MPS representation:
\begin{equation}
S_j^z\ket{S_0}=\sum_{\mu_1,\mu_2,\cdots,\mu_L}\textnormal{Tr}\left[A^{(\mu_1)}\cdots A^{(\mu_{j-1})}B^{(\mu_{j})}A^{(\mu_{j+1})} \cdots A^{(\mu_L)}\right]\ket{\mu_1,\mu_2,\cdots,\mu_L},
\end{equation}
where $B^{(\pm)}=\pm A^{(\pm)},B^{(0)}=0$. Noticing that $B^{(\mu)}=[C,A^{(\mu)}]$ with $C=-\frac{1}{2}\sigma^z$, the local action of $S_j^z$ is given by  
\begin{equation}
    S_j^z\ket{AAA}_{j-1,j,j+1}=\ket{(ACA)A}_{j-1,j,j+1}-\ket{A(ACA)}_{j-1,j,j+1}.
\end{equation}
where uncontracted indices are omitted. The matrices $A^{(\mu)}CA^{(\nu)}$ can be rewritten into a bilinear form of $A$ matrices as $A^{(\mu)}CA^{(\nu)}=\sum_{\mu',\nu'}c_{\mu,\nu;\mu',\nu'}A^{(\mu')}A^{(\nu')}$, where explicit values of the coefficients $c_{\mu,\nu;\mu',\nu'}$ are not shown here. Then we can construct the operator $O_{j,k}$ with the same coefficients as
\begin{equation}
 O_{j,k}=\sum_{\mu,\nu;\mu',\nu'}c_{\mu,\nu;\mu',\nu'}(\ket{\mu,\nu}\bra{\mu',\nu'})_{j,k}
\end{equation}
Therefore, the non-Hermitian operator $(S_j^z-O_{j-1,j}+O_{j,j+1})$ annihilates the three-local states $\ket{AAA}$ of $\ket{S_0}$. This technique of decomposition works for other scarred models with zero-energy MPS eigenstates as well (e.g. $E=0$ scar states in the PXP model  \cite{lin2019exact}). Notice that in this case the non-Hermitian deformation enlarges the support of the original local operator, which is vastly different with those in the main text and the DMI Hamiltonian.

We give a brief comment on the type-II symmetric Hamiltonian for the AKLT scar tower.  The Hamiltonian $H_{\text{AKLT}}-S^z_{\text{tot}}$ can be decomposed to local terms as
\begin{equation}
    H_{\text{AKLT}}-S^z_{\text{tot}}=\sum_j[T_{j,j+1}^{S=2}-\frac{1}{2}(S_j^z+S_{j+1}^z)]=\sum_j[2(T_{j,j+1}^{S=2,m=-2}+T_{j,j+1}^{S=2,m=-1})+\frac{1}{2}(\ket{0,1}\bra{1,0}-\ket{0,-1}\bra{-1,0}+\text{H.c.})_{j,j+1}].
\end{equation}
Here the projectors $T_{j,j+1}^{S=2,m=-2}$ and $T_{j,j+1}^{S=2,m=-1}$ directly annihilate the scar tower, so we should find the non-Hermitian deformation for the remaining terms $h_{j,j+1}=(\ket{0,1}\bra{1,0}-\ket{0,-1}\bra{-1,0}+\text{H.c.})_{j,j+1}$. From the above analysis on the AKLT ground state, we infer that the deformation of $h_{j,j+1}$ may involve some three-local operators $\tilde{O}_{j,k,l}$, which results in the four-local non-Hermitian operator $h_{j,j+1}-\tilde{O}_{j-1,j,j+1}+\tilde{O}_{j,j+1,j+2}$ annihilating the scar tower.

\subsection{Criterion for local non-Hermitian annihilators}\label{sec:criterion}
In the previous subsections, we have shown that several scarred Hamiltonians can be decomposed to local non-Hermitian annihilators, hosting the scar states in their common null space. Now, we shift our focus to the inverse problem: How to efficiently find Hermitian Hamiltonians annihilating a given tower of scar states? 
To address the problem, we propose a systematic method, particularly focusing on searching for those extensive local Hermitian annihilators which are not easily approached by previous methods, i.e. the projector embedding. For the sake of simplicity, we will study those annihilators as the translational-invariant summation of two-local operators under PBC.

Assuming that local projectors $P_{j,j+1}$ annihilating a tower of scar states have already been obtained, e.g. through the compressed MPS technique \cite{Shibata2020Onsager,Chattopadhyay2020quantum,Mark2020Unified,Wang2023Embedding}. 
We can then construct local annihilators (not necessarily Hermitian) as $\tilde{h}_{j,j+1}$, subjected to two constraints: 
\begin{equation}
    \tilde{h}_{j,j+1}(1-P_{j,j+1})=0,\qquad P_{j,j+1}\tilde{h}_{j,j+1}P_{j,j+1}=P_{j,j+1}\tilde{h}_{j,j+1}=0.
\end{equation}
The first constraint ensures that $\tilde{h}_{j,j+1}$ locally annihilates the scar tower, while the second excludes those already known Hamiltonians within the framework of Shiraishi-Mori embedding. Under the two constraints, the local annihilator admits the form $ \tilde{h}_{j,j+1}=(1-P_{j,j+1})\tilde{h}_{j,j+1}P_{j,j+1}$, with $\text{ker}(P_{j,j+1})\times \text{ker}(1-P_{j,j+1})$ non-zero matrix elements to be determined. The summation of those local annihilators under PBC is given by
\begin{equation}
    \tilde{H}_L=\sum_{j=1}^L \tilde{h}_{j,j+1},\qquad L+1\equiv 1,
\end{equation}
which is demanded to be Hermitian for any system size $L$.

We find that it suffices to impose the Hermitian conditions only for $\tilde{H}_2$ and $\tilde{H}_3$, rather than arbitrary $L$. Our proof is inductive: Suppose that for a certain integer $L\ge 3$, $\tilde{H}_{L'}$ has been proved to be Hermitian for all $2\le L'\le L$. We can rewrite $\tilde{H}_{L+1}$ as
\begin{align}
    \tilde{H}_{L+1}=&\tilde{H}_{L}-\tilde{h}_{L,1}+\tilde{h}_{L,L+1}+\tilde{h}_{L+1,1}\nonumber\\
    =&\tilde{H}_{L}-(\tilde{h}_{L,1}+\tilde{h}_{1,L})+(\tilde{h}_{1,L}+\tilde{h}_{L,L+1}+\tilde{h}_{L+1,1})
\end{align}
Notice that $(\tilde{h}_{L,1}+\tilde{h}_{1,L})$ is isomorphic to $\tilde{H}_2$, while $(\tilde{h}_{1,L}+\tilde{h}_{L,L+1}+\tilde{h}_{L+1,1})$ is isomorphic to $\tilde{H}_3$. Hence, $\tilde{H}_{L+1}$ is Hermitian, and we have completed the proof.

Therefore, to exhaust all solutions $\tilde{h}_{j,j+1}$ fulfilling the Hermitian conditions, we only need to solve two equations as the criterion: $\tilde{H}_{2(3)}=\tilde{H}_{2(3)}^\dagger$. These equations for local operators can be quickly solved analytically or numerically. In the following, we apply the method to the ferromagnetic Dicke states to solve their extensive local annihilators, and show that the DMI Hamiltonians (Sec. \ref{sec:DMI}) are the only non-trivial solutions.

Two-local projectors for the Dicke states project two adjacent spin-$1/2$'s onto the spin-$0$ singlet, which reads $P_{j,j+1}=(\ket{S=0}\bra{S=0})_{j,j+1}$, where $S$ denotes the total spin of two spins on site $j$ and $j+1$. Under the two constraints, we propose an ansatz for the local annihilators as:
\begin{equation}
    \tilde{h}_{j,j+1}=\sum_{m=0,1,-1}a_m (\ket{S=1,m}\bra{S=0})_{j,j+1},
\end{equation}
where $m$ denotes the total spin-$z$ component of two spins on site $j$ and $j+1$, and $a_m$ are three complex-valued coefficients to be determined. The $L=2$ PBC Hamiltonian is zero, since $\ket{S=0}_{1,2}$ is anti-symmetric by exchanging two spins, while $\ket{S=1,m}_{1,2}$ are symmetric, and thus $\tilde{h}_{1,2}+\tilde{h}_{2,1}=0$. Solving the Hermitian condition for the $L=3$ PBC Hamiltonian results in $a_0=-a_0^*,a_1=a_{-1}^*$, which correspond to three linearly independent solutions:
\begin{enumerate}
    \item $a_0=i, a_{\pm 1}=0$: $\tilde{h}_{j,j+1}=i(\ket{S=1,0}\bra{S=0})_{j,j+1}$, sums to $H_{z-\text{DMI}}$;
    \item $a_{\pm 1}=1,a_0=0$: $\tilde{h}_{j,j+1}=[(\ket{S=1,1}+\ket{S=1,-1})\bra{S=0}]_{j,j+1}$, sums to $H_{x-\text{DMI}}$;
    \item $a_{\pm 1}=\pm i,a_0=0$: $\tilde{h}_{j,j+1}=i[(\ket{S=1,1}-\ket{S=1,-1})\bra{S=0}]_{j,j+1}$, sums to $H_{y-\text{DMI}}$.
\end{enumerate}
The DMI Hamiltonian $H_{\alpha-\text{DMI}}$ with arbitrary $\hat{\alpha}$ can be obtained by linear superpositions of the three solutions, which annihilates the tower of Dicke states.

\section{Bipartite entanglement entropy}
In this section, we obtain the bipartite entanglement entropy of scar states in the $N=3$ model. Results for larger $N$ follow a similar route but the derivation will be much more sophisticated. To calculate the bipartite entanglement entropy of a scar state $\ket{S_{m_1,m_2,m_3}}$ with $m_1+m_2+m_3=L$ between two subsystems $A$ and $\bar{A}$, we need to trace out $\bar{A}$ and get the reduced density matrix on the subsystem $A$: $ \rho_A=\text{Tr}_{\bar{A}}[\ket{S_{m_1,m_2,m_3}}\bra{S_{m_1,m_2,m_3}}]$, then the entanglement entropy (\text{EE}) is given by $\text{EE}=-\text{Tr}[\rho_A\ln(\rho_A)]=-\sum \lambda_A\ln(\lambda_A)$, where $\lambda_A$ are eigenvalues of $\rho_A$. To extract $\lambda_A$, we can apply Schmidt decomposition on the target state so that $\lambda_A$ are given by Schmidt coefficients. To this end, we introduce two raising operators 
\begin{equation}
    Q^\dagger_{\pm}=\sum_{j=1}^L\exp(i(j-1)\gamma)(\ket{\pm 1}\bra{0})_j,
\end{equation}
then the scar state can be generated from the ``vacuum" $\ket{\Omega}=\ket{S_{L,0,0}}$:
\begin{equation}
    \ket{S_{m_1,m_2,m_3}}=\frac{1}{\sqrt{N(m_2,m_3;L)}}(Q^\dagger_{+})^{m_2} (Q^\dagger_{-})^{m_3} \ket{\Omega}.
\end{equation}
The normalization factor $N(m_2,m_3;L)=(m_2)!(m_3)!(L)!/(L-m_2-m_3)!$ is given by counting the number of different configurations (computational basis states) in $(Q^\dagger_{+})^{m_2} (Q^\dagger_{-})^{m_3} \ket{\Omega}$, and the multiplicity of each basis state  \cite{Popkov2005Log}. In order for the Schmidt decomposition, raising operators are divided as $Q^\dagger_{\pm}=Q^\dagger_{A\pm}+Q^\dagger_{\bar{A}\pm}$, where for each part the summation of $j$ only covers the corresponding subsystem. Consequently, we have the bipartition:
\begin{align}
    \ket{S_{m_1,m_2,m_3}}=&\frac{1}{\sqrt{N(m_2,m_3;L)}}(Q_{A+}^\dagger+Q_{\bar{A}+}^\dagger)^{m_2}(Q_{A-}^\dagger+Q_{\bar{A}-}^\dagger)^{m_3}\ket{\Omega}\nonumber\\
    =&\frac{1}{\sqrt{N(m_2,m_3;L)}}\sum_{k_2}\sum_{k_3}\binom{m_2}{k_2}\binom{m_3}{k_3}(Q_{A+}^\dagger)^{k_2}(Q_{A-}^\dagger)^{k_3}(Q_{\bar{A}+}^\dagger)^{m_2-k_2}(Q_{\bar{A}-}^\dagger)^{m_3-k_3}\ket{\Omega}\nonumber\\
    =&\sum_{k_2=\text{max}[0,m_2-(L-L_A)]}^{\text{min}(m_2,L_A)}\sum_{k_3=\text{max}[0,(m_2+m_3)-(L-L_A)-k_2)]}^{\text{min}(m_3,L_A-k_2)}\sqrt{\frac{N(k_2,k_3;L_A)N(m_2-k_2,m_3-k_3;L-L_A)}{N(m_2,m_3;L)}}\nonumber\\
    &\binom{m_2}{k_2}\binom{m_3}{k_3}\ket{S_{L_A-k_2-k_3,k_2,k_3}}\otimes \ket{S_{L+m_2+m_3-L_A-k_2-k_3,m_2-k_2,m_3-k_3}},
\end{align}
and eigenvalues of the reduced density matrix can be extracted immediately:
\begin{equation}
\lambda_{k_2,k_3}=\frac{N(k_2,k_3;L_A)N(m_2-k_2,m_3-k_3;L-L_A)}{N(m_2,m_3;L)}\binom{m_2}{k_2}^2\binom{m_3}{k_3}^2.
\end{equation}

\begin{figure}
\hspace*{-0.6\textwidth}
\includegraphics[width=0.6\linewidth]{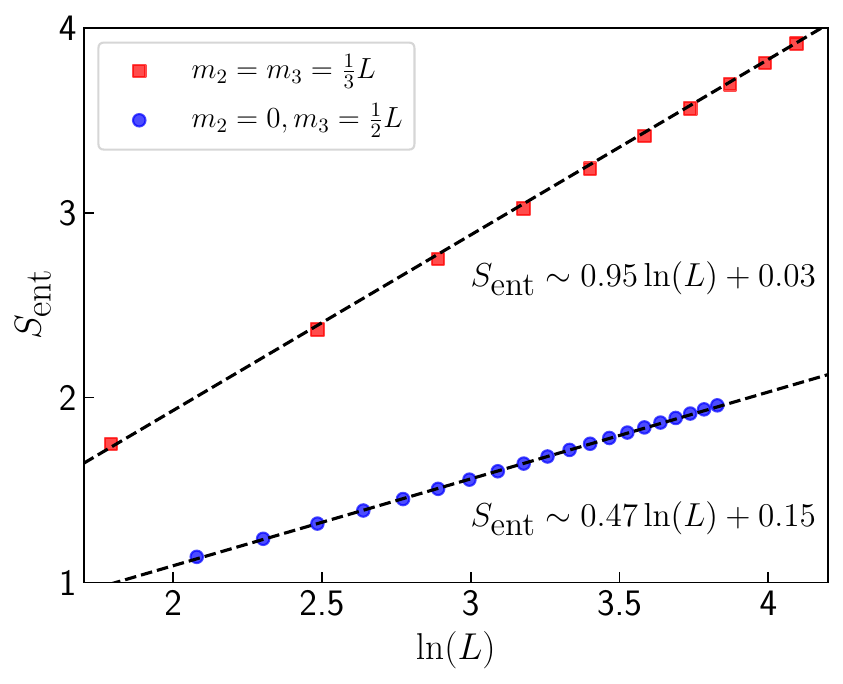} 
\caption{Bipartite entanglement entropy $S_\text{ent}$ scaling ($L_A=L/2$) of the scar states in the $N=3$ model. Red squares: $m_2=m_3=L/3$, blue dots: $m_2=0,m_3=L/2$. Both $S_\text{ent}$ grow logarithmically with the system size. }
\label{Scaling}
\end{figure}

Since $\sum_{k_2,k_3}\lambda_{k_2,k_3}=1$, these eigenvalues can be regarded as the distribution probability of two random variables $k_2$ and $k_3$. Further, it admits a form of $\lambda_{k_2,k_3}=p(k_3)p(k_2|k_3)$, where
\begin{equation}
    p(k_3)=\sum_{k_2}\lambda_{k_2,k_3}=\binom{L}{L_A}^{-1}\binom{m_3}{k_3}\binom{L-m_3}{L_A-k_3},
\end{equation} 
and 
\begin{equation}
    p(k_2|k_3)=\lambda_{k_2,k_3}/p(k_3)=\binom{L-m_3}{L_A-k_3}^{-1}\binom{m_2}{k_2}\binom{L-m_2-m_3}{L_A-k_2-k_3}
\end{equation}
is the conditional probability of $k_2$ conditioned on $k_3$. Given the expression, we find the entropy can be decomposed as
\begin{equation}\label{eq:sumcomb}
    S_\text{ent}=-\sum_{k_2,k_3}\lambda_{k_2,k_3}\ln(\lambda_{k_2,k_3})=-\sum_{k_3}p(k_3)\ln(p(k_3))-\sum_{k_3}p(k_3)\sum_{k_2}p(k_2|k_3)\ln(p(k_2|k_3)).
\end{equation}
Consider the states with large $L$ and finite $m_{1,2,3}/L$ (i.e. finite densities of excitations), the entanglement entropy can be evaluated through the standard method of approximations \cite{Popkov2005Log,Schecter2019Weak}: The first terms give
\begin{equation}
    -\sum_{k_3}p(k_3)\ln(p(k_3))\approx\frac{1}{2}\ln(\frac{L_A(L-L_A)}{L})+\frac{1}{2}\ln(2\pi e \frac{m_3(L-m_3)}{L^2})\approx \frac{1}{2}\ln(L)+\mathcal{O}(1),
\end{equation}
and the second is
\begin{align}
    -\sum_{k_3}p(k_3)\sum_{k_2}p(k_2|k_3)\ln(p(k_2|k_3))&\approx\sum_{k_3}\frac{1}{2}p(k_3)(\ln(\frac{(L_A-k_3)(L-L_A-m_3+k_3)}{L-m_3})+\ln(2\pi e \frac{m_2(L-m_2-m_3)}{(L-m_3)^2}))\nonumber\\
    &\approx \frac{1}{2}\ln(L)+\mathcal{O}(1).
\end{align}
Therefore, conceptually two kinds of excitations $\ket{\pm 1}$ contribute equally to the leading order of the entanglement entropy, and they sum to $\sim\ln(L)$. In Fig. \ref{Scaling}, we numerically calculate the entanglement entropy with $L_A=L/2$ for the scar state $\ket{S_{L/3,L/3,L/3}}$ through Eq. \eqref{eq:sumcomb}. We find that $S_\text{ent}$ grows logarithmically with the system size $L$, with a coefficient approximating one as predicted. As a comparison, we also display the entanglement entropy scaling for $\ket{S_{L/2,0,L/2}}$, which exhibits logarithmic growth with the coefficient $\sim 0.5$, since only $\ket{-1}$ excitations contribute to the entanglement.

\bibliographystyle{apsrev4-1-title}
\bibliography{DengQAIGroup,QMBS,Heran_Helix}